\documentclass[a4paper,11pt]{article}
\usepackage{pos}

\title{Decay modeling in 4top production}

\author*[a]{Manal Alsairafi}

\affiliation[a]{Institute for Theoretical Particle Physics and Cosmology, RWTH Aachen University, D-52056 Aachen, Germany}

\emailAdd{manal.alsairafi@rwth-aachen.de}

\abstract{We present NLO QCD predictions for the $pp\to t\bar{t}t\bar{t}+X$ process in the 4 lepton decay channel at the integrated and differential level for the LHC centre-of-mass energy of $\sqrt{s}=13.6$~TeV. Specifically, we compare the fixed order NLO QCD calculations to the parton-shower matched results obtained with \textsc{MC@NLO} and \textsc{Powheg} matching. In the former, NLO QCD corrections are consistently included in the production and decay of the four top quarks, preserving all spin correlations. In the latter, higher order effects in top-quark decays with approximate spin correlations are simulated in the parton showers, using Pythia. Additionally the impact of including matrix element corrections in the top quark decays in the parton-shower matched predictions is analysed.}

\FullConference{14th Edition of the Large Hadron Collider Physics (LHCP2026)\\
18-22 May 2026\\
Paris, France\\}

\begin{document}

\renewcommand{\hookAfterAbstract}{\par\bigskip
P3H-26-079, TTK-26-39}

\maketitle

\section{Introduction}
Although four-top production has a very small cross section at the LHC, the process is of considerable interest for several reasons. First, taking into account top-quark decays in the process, we are dealing with a $2\to12$ process with four $b$-quarks in the final state. This provides a good avenue to test the standard model, specifically QCD corrections and Parton Shower (PS) effects. 
Second, due to the large top-quark mass and the process' rarity, it is sensitive to new physics beyond the standard model.
The process has been recently observed by both the ATLAS and CMS collaborations at the LHC \cite{ATLAS:2023ajo,CMS:2023ftu}. Three signal regions were considered, the four, three and two same-sign lepton channels.
On the theory aspect, the first calculations of the process at NLO in QCD with stable top quarks were provided in \cite{Bevilacqua:2012em,Maltoni:2015ena}. Later, EW corrections were additionally considered in \cite{Frederix:2017wme}. However, the most precise prediction for stable four-top production is obtained at next-to-leading logarithmic accuracy in \cite{vanBeekveld:2022hty,vanBeekveld:2025ghw}.
Nevertheless, considering the decays of the top quarks is important for the study of the process in the fiducial phase-space. Predictions at NLO+PS accuracy have been provided in \cite{Jezo:2021smh}, particularly, the $1\ell+jets$ channel was considered through \textsc{Powheg} matching. Moreover, results at the fixed-order level including top-quark decays in the $4\ell$ and $3\ell$ channels were obtained in \cite{Dimitrakopoulos:2024qib,Dimitrakopoulos:2024yjm}. Here, the Narrow Width Approximation (NWA) is taken into account with different treatments. Finally, the two approaches in the $4\ell$ are compared in \cite{Alsairafi:2025rjd}. This contribution highlights and summarises some of the results in \cite{Alsairafi:2025rjd}.

\section{Description of the Calculation}
We study the modelling of the top-quark decays in the four-top process via two methods, with the setup as described in \cite{Alsairafi:2025rjd}. 
First, we calculate the process at the fixed-order level at NLO QCD in the NWA and additionally considering the perturbative expansion of the top-quark decay width, thus referring to the treatment as $\rm NWA_{exp}$. This treatment, as compared to the full NWA treatment is thoroughly described in \cite{Dimitrakopoulos:2024qib}. This calculation is preformed in the \textsc{Helac-Nlo} framework \cite{Bevilacqua:2011xh}.
Second, we generate the results matched to parton showers (NLO+PS) with both \textsc{MC@NLO} matching, via \textsc{MadGraph5${}_{-}$aMC@NLO} \cite{Frixione:2002ik}, and \textsc{Powheg} matching, within the \textsc{Powheg-Res} framework \cite{Nason:2004rx,Frixione:2007vw,Alioli:2010xd}. \textsc{Pythia8.3} is utilised to shower the generated events  \cite{Bierlich:2022pfr}, which are then analysed through \textsc{Rivet} \cite{Buckley:2010ar, Bierlich:2019rhm}. The latter is interfaced to the \textsc{HepMC3} \cite{Buckley:2019xhk} and \textsc{FastJet}  \cite{Cacciari:2005hq, Cacciari:2011ma} libraries. Additionally, the effects of Matrix-Element Corrections (MECs) to the NLO+PS events are studied. The MECs are applied according to \cite{Frixione:2023hwz}.

We now describe the differences between the two modelling treatments.
First, in the fixed-order calculation, higher order corrections to both the production and decay of the four top quarks are included, consequently, NLO accurate spin correlations are preserved. On the other hand, in the NLO+PS calculation, the top quarks are produced with NLO accuracy, but their decays are included at LO accuracy. Thus, spin correlations are only LO accurate. The additional parton shower in this calculation approximates higher order corrections with collinear and soft-collinear emissions. This approximation is further improved in the decays of the top quarks by implementing the MECs.

Second, in the fixed-order NLO calculation, we generate at most five final state $b$-jets, four of which are due to the top-quark decays. In contrast, the NLO+PS calculation predicts a larger number of final-state $b$-jets due to the additional radiation generated by the parton shower. These additional $b$-jets may be harder than those originating from top-quark decays. Thus, using Monte Carlo truth information, we can identify the $b$-jets due to the top-quark decays and apply our selection cuts accordingly.

Finally, we use the dynamical scale choice $\mu_R=\mu_F=E_T/4$, where $E_T$ is the sum of transverse masses of the (anti-)top quarks.
This setting in the fixed-order calculation is based on the final state jets and leptons in the fiducial phase space, which are used to reconstruct the top-quark kinematics. However, the scale choice in the NLO+PS results is based directly on the top quark kinematics prior to their decay. Additionally, the transverse mass of the additional particle due to the real emission correction is included in the sum of $E_T$.
\begingroup
\begin{table}[t!]
\centering
\begin{tabular}{ccccccc}
  \hline\hline
\textsc{Theoretical Approach} & $\sigma$ [ab] & $+\delta^{\rm scale}$&$-\delta^{\rm scale}$& +$\delta^{\rm match}$ & -$\delta^{\rm match}$ & $\sigma/\sigma_{\rm NWA_{exp}}$ \\
   \hline\hline
\multicolumn{7}{c}{\textsc{Fixed-Order NLO}} \\
   \hline\hline
${\rm NWA_{exp}}$ & 5.170(3) & +12\% & -20\% & - & - & 1.0000 \\
   \hline\hline
\multicolumn{7}{c}{\textsc{NLO+PS without $b$-Jet Identification}} \\
   \hline\hline
${\rm MC@NLO+PS}$ & 4.891(8) & +15\% & -21\% & +1.1\% & -0.7\% & 0.9460 \\
${\rm MC@NLO+PS\oplus MEC}$ & 5.098(8) & +15\% & -21\% & +0.7\% & -0.5\% & 0.9861 \\
\hline
${\rm POWHEG+PS}$  & 4.820(6) & +15\% & -21\% & +0.0\% & -0.5\% & 0.9323\\
${\rm POWHEG+PS\oplus MEC}$& 5.024(6) & +15\% & -21\% & +0.0\% & -0.5\% & 0.9718  \\
   \hline\hline
\multicolumn{7}{c}{\textsc{NLO+PS with $b$-Jet Identification}} \\
   \hline\hline
${\rm MC@NLO+PS}$ & 4.708(7) & +15\% & -21\% & +0.8\% & -0.4\% & 0.9106 \\
${\rm MC@NLO+PS\oplus MEC}$ & 4.923(8) & +15\% & -21\% & +0.5\% & -0.5\% & 0.9522 \\
\hline
${\rm POWHEG+PS}$  & 4.656(6)   & +15\% & -21\% & +0.1\% & -0.2\% & 0.9006  \\
${\rm POWHEG+PS\oplus MEC}$& 4.865(6) & +15\% & -21\% & +0.0\% & -0.2\% & 0.9410  \\
   \hline\hline
\end{tabular}
    \caption{\textit{Integrated fiducial cross-section predictions
    at NLO in QCD. The fixed-order
    results (denoted by $\rm NWA_{exp}$) and NLO+PS results are presented. The parton-shower matched results are
    displayed with and without the $b$-jet identification as well as with and without MECs. The scale uncertainties, $\delta^{\rm scale}$, and parton-shower matching uncertainties, $\delta^{\rm match}$, are also shown. These results are taken from \cite{Alsairafi:2025rjd}.}}
    \label{table: 4lepton integrated}
\end{table}
\endgroup
\begin{figure}[t]
    \centering
    \includegraphics[width=0.75\linewidth]{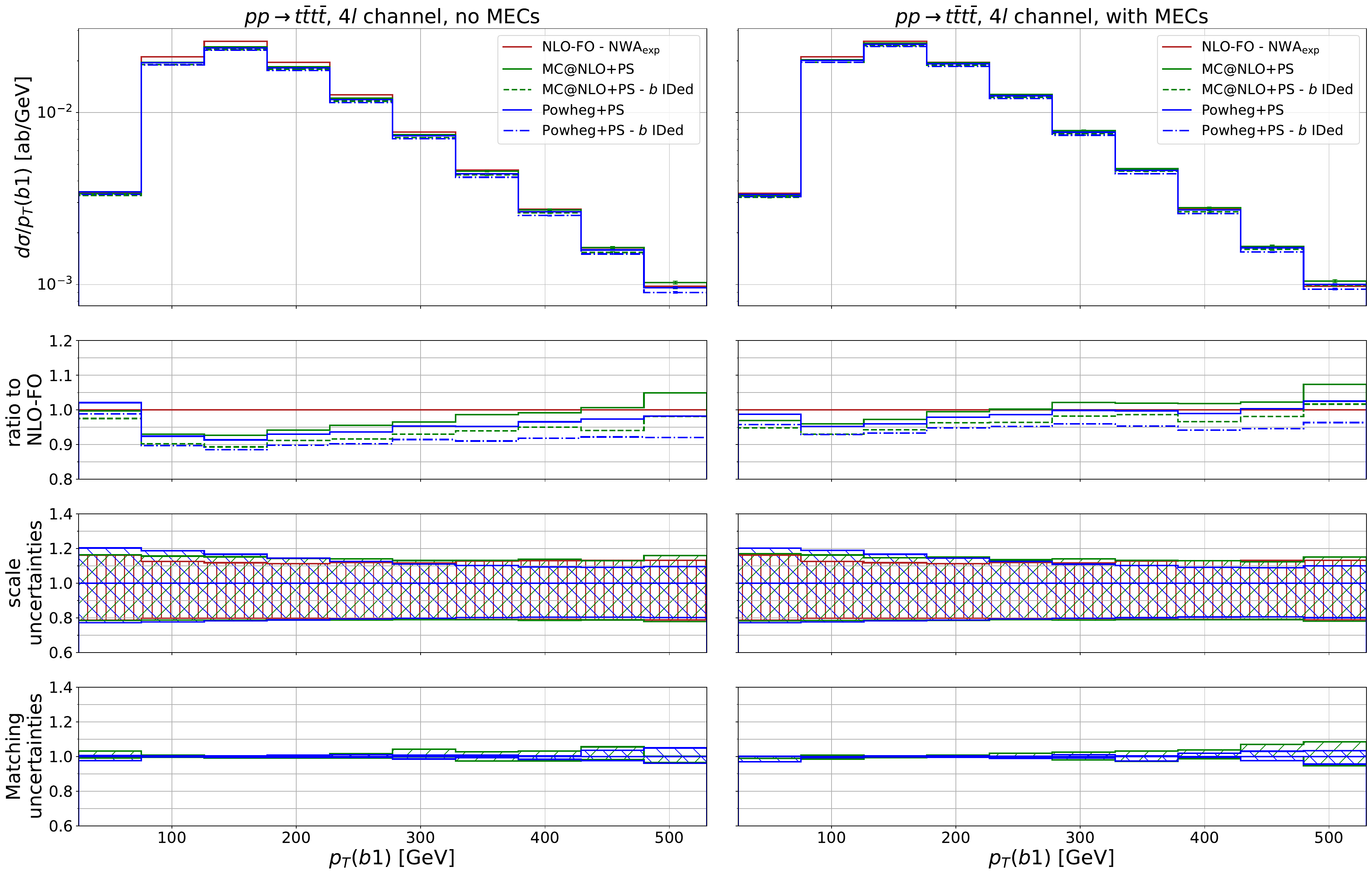}
    \caption{\textit{Differential cross-section distributions of the process as a function of the transverse momentum of the hardest $b$-jet, $p_T(b_1)$. The NLO+PS results are given without (left) and with the MEC (right). Predictions without (solid line) and with (dashed line) the $b$-jet identification are also shown. The second panel provides the ratio to the fixed-order prediction. Theoretical scale and matching uncertainties are displayed in the bottom two panels. Figure taken from \cite{Alsairafi:2025rjd}.}}
    \label{fig:pT(b1)}
\end{figure}

\begin{figure}[t]
    \centering
    \includegraphics[width=0.75\linewidth]{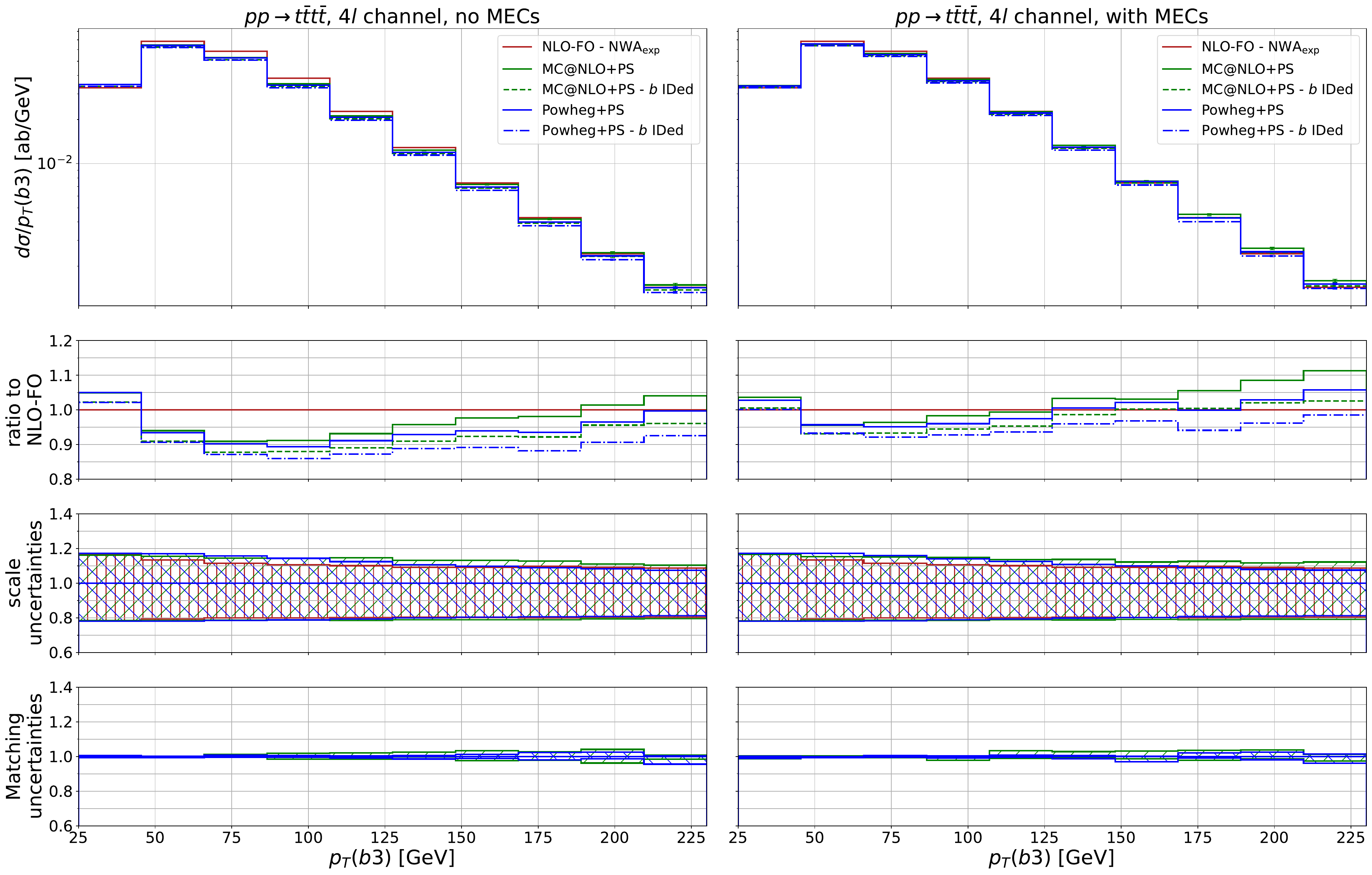}
    \caption{\textit{Same as figure~\ref{fig:pT(b1)} but for the transverse momentum of the third hardest $b$-jet. Figure taken from \cite{Alsairafi:2025rjd}.}}
    \label{fig:pT(b3)}
\end{figure}

\begin{figure}[t]
    \centering
    \includegraphics[width=0.75\linewidth]{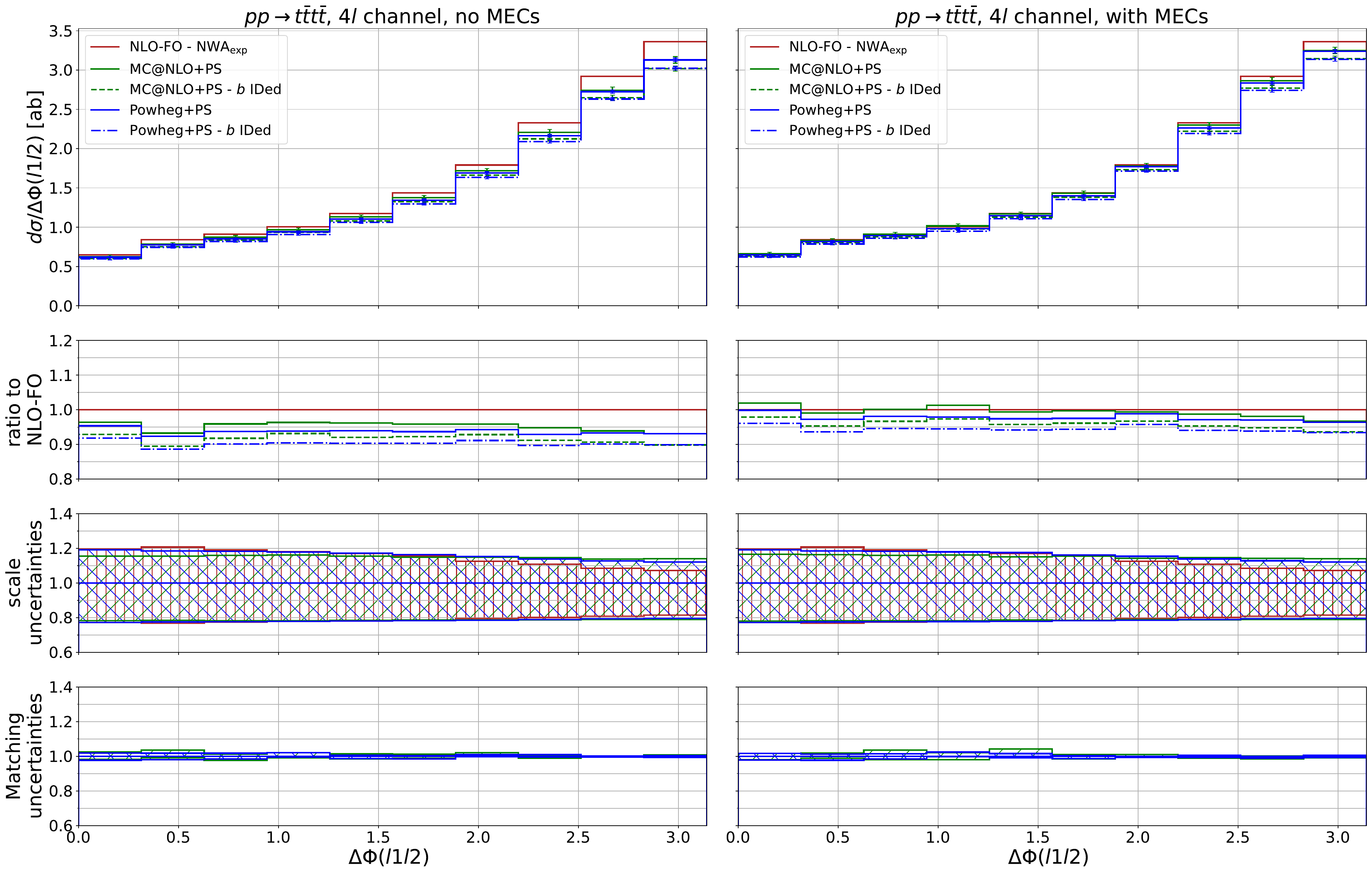}
    \caption{\textit{Same as figure~\ref{fig:pT(b1)} but for the angular distribution of the hardest two leptons. Figure taken from \cite{Alsairafi:2025rjd}.}}
    \label{fig:dphi}
\end{figure}

\section{Results}
We start by presenting the integrated fiducial cross-sections, shown in table~\ref{table: 4lepton integrated}. We note that all results are compatible within the theoretical uncertainties. These uncertainties are similar to the projected HL-LHC statistical uncertainties, which are about 18\%-16\%. We also note that MECs increase the corresponding cross-sections by around 4\%, increasing the agreement between the NLO+PS and fixed-order predictions. Furthermore, identifying $b$-jets originating from top-quark decays reduces the cross-section as the four hardest $b$-jets which pass the selection cuts may not always be those from the hard process.

We now move on to differential distributions, starting with the transverse momentum of the hardest $b$-jet in figure~\ref{fig:pT(b1)}. We notice an improvement in the shape of the NLO+PS distributions after applying MECs. 
Second, we study the transverse momentum of the third hardest $b$-jet in figure~\ref{fig:pT(b3)}. Here, we observe a similar trend as for $p_T(b_1)$, where MECs improve the NLO+PS results.
We additionally find, specifically for the \textsc{MC@NLO} prediction, that $b$-jet identification has an impact. This can be seen by the deviation in the shape of the curves between the results with and without the identification.
Although we established that the compatibility of the results improves with the application of MECs and $b$-jet identification, we should note that regardless, all results agree within theoretical uncertainties.

Lastly, we look at the angular distribution $\Delta\Phi(\ell_1\ell_2)$, depicted in figure~\ref{fig:dphi}, which is an observable sensitive to spin correlations. We find that all results agree in shape and within theoretical uncertainties. Applying MECs here shifts the normalisation of the distribution on a similar degree as on the integrated level.

\section{Conclusions}
We have performed a comparative study between fixed-order and NLO+PS predictions for the four-top quark production and decay in the fully leptonic channel. All results are in agreement within the theoretical uncertainties. Nevertheless, we note an improvement in the agreement of the various central predictions with the application of matrix element corrections on the NLO+PS calculation. Thus we establish that MECs partially recover higher order corrections in the top-quark decays. We analysed specifically three distributions. 
The transverse momenta of the hardest and third hardest $b$-jets demonstrate this, where the shape of the NLO+PS distribution is impacted. 
We further examined the spin-correlation-sensitive observable, $\Delta\Phi(\ell_1\ell_2)$, and observed that the addition of MECs to the NLO+PS predictions also leads to a better agreement with the fixed order results, however through a shift in the normalisation of the distribution.
\newpage
\subsection*{Acknowledgements}

This work was supported by the Deutsche Forschungsgemeinschaft (DFG) under grant 396021762 - TRR 257: \textit{Particle Physics Phenomenology after the Higgs Discovery}, and grant 400140256 - GRK 2497: \textit{The Physics of the Heaviest Particles at the LHC.}

\end{document}